\documentclass[aps,prd,a4paper,superscriptaddress,nofootinbib,
showpacs,twocolumn,showkeys,amsfonts,amssymb,amsmath]{revtex4}

\usepackage{amssymb,latexsym}
\usepackage{amsmath, amsthm}
\usepackage{amscd}
\usepackage{times}
\usepackage{epsfig}
\usepackage{psfrag}
\usepackage{graphicx}
\usepackage{overpic}
\usepackage{booktabs,tabularx}

\newcommand{\be}{\begin{equation}}
\newcommand{\ee}{\end{equation}}
\newcommand{\bea}{\begin{eqnarray}}
\newcommand{\eea}{\end{eqnarray}}

\begin{document}

\title{Observational properties of rigidly rotating dust configurations}

\author{Batyr~Ilyas}
\email{batyr.ilyas@nu.edu.kz}
\affiliation{Department of Physics,
Nazarbayev University, 53 Kabanbay Batyr, 010000 Astana, Kazakhstan}

\author{Jinye~Yang}
\email{yangjy14@fudan.edu.cn}
\affiliation{Center for Field Theory and Particle Physics and Department of Physics,
Fudan University, 220 Handan Road, 200433 Shanghai, China}

\author{Daniele~Malafarina}
\email{daniele.malafarina@nu.edu.kz}
\affiliation{Department of Physics,
Nazarbayev University, 53 Kabanbay Batyr, 010000 Astana, Kazakhstan}

\author{Cosimo~Bambi}
\email{bambi@fudan.edu.cn}
\affiliation{Center for Field Theory and Particle Physics and Department of Physics,
Fudan University, 220 Handan Road, 200433 Shanghai, China}
\affiliation{
Theoretical Astrophysics, Eberhard-Karls Universit\"at T\"ubingen, Auf der Morgenstelle 10, 72076 T\"ubingen, Germany}


\begin{abstract}
We study the observational properties of a class of exact solutions of Einstein's field equations describing stationary, axially symmetric, rigidly rotating dust (i.e. non interacting particles). We ask the question whether such solutions can describe astrophysical rotating dark matter clouds near the center of galaxies and we probe the possibility that they may constitute an alternative to supermassive black holes at the center of galaxies.
We show that light emission from accretion disks made of ordinary baryonic matter in this space-time has several differences with respect to the emission of light from similar accretion disks around black holes.
The shape of the iron K$\alpha$ line in the reflection spectrum of accretion disks can potentially distinguish this class of solution from the Kerr metric, but this may not be possible with current X-ray missions.
\end{abstract}

\maketitle



\section{Introduction}\label{intro}

We know that supermassive compact objects exist at the center of galaxies~\cite{revsm}. It is usually believed that they must be black holes even though at present we do not have any conclusive evidence about their nature. For this reason it is worth investigating the possibility that these sources may not be black holes and if there are possible observational tests that can distinguish between black holes and other, more exotic, sources of the gravitational field~\cite{review,test1,test2,test3}. The general theory of relativity allows for several exact solutions that do not describe black holes and that could be used to model gravitating compact objects. Indeed in the last few years there has been great interest in the study of the motion of test particles on accretion disks around exotic compact objects and naked singularities and in the possible ways to distinguish such space-times from black holes (see for example~\cite{obs,obs2,obs3}).

The study of the disk's reflection spectrum is a promising tool to observationally test the space-time metric around these supermassive objects~\cite{i1,i2,i3,i4}. The most prominent feature is the iron K$\alpha$ line, which is very narrow in the rest frame of the gas and is instead broad and skewed in the observed spectrum of a black hole candidate due to relativistic effects (gravitational redshift, Doppler boosting, light bending) occurring in the strong gravity region. In the presence of high quality data and with the correct astrophysical model, this approach can provide stringent constraints on the nature astrophysical black hole candidates.

In Ref.~\cite{BM}, two of us studied the shape of the iron K$\alpha$ line coming from accretion disks inside low density matter clouds with a central singularity and compared with the same line produced by accretion disks around black holes. In the present article, we focus on rotating non vacuum solutions as possible candidates for supermassive compact objects. This is an ideal natural continuation of the work presented in Ref.~\cite{BM}, as here we consider an axially symmetric matter source with non vanishing angular momentum.

The model is made of three elements: (i) As the matter source that determines the geometry we consider a cloud of rigidly rotating dust (i.e. non interacting particles). (ii) Due to the divergence of the energy density at the center we consider a cutoff of the metric at a minimum radius $r_0$ and assume the existence of an exotic compact object at the center. (iii) In order to test the observable features of the space-time we consider a luminous accretion disk made of ordinary matter in the equatorial plane and we compare its features with the features of similar disks around black holes.

Stationary, axially symmetric space-times are of great importance in astrophysics as they may be connected to the origin of extragalactic jets. For example, in Ref.~\cite{Bonnor}, it was shown that rigidly rotating dust configurations present a density gradient parallel to the axis, which is missing in the corresponding Newtonian case. The argument, put forward by Opher, Santos and Wang in Ref.~\cite{OSW} is that the space-time geometry of a rotating dust cylinder, such as that presented in Ref.~\cite{Bonnor2}, can contribute to the collimating effects observed in extragalactic jets. Unfortunately, when considering rotating fluids in general relativity (GR), there are several pathologies that may arise, such as singularities, non asymptotic flatness and negative densities, and these may hinder our understanding of the physical significance of the source.

 In the present article, we study accretion disks in the interior of a fluid body made of weakly interacting particles. One can think of such a pressureless fluid which does not interact with ordinary particles as describing the central region of a dark matter cloud. To this aim, we will focus on an exact solution describing a rigidly rotating dust cloud that was first discussed by Bonnor in Ref.~\cite{Bonnor}. The Bonnor solution is a member of the family of stationary, axially symmetric space-times that was derived by Winicour in Ref.~\cite{winicour}. This solution is particularly appealing because it is asymptotically flat and its energy density is everywhere positive, so it is physically viable, except for its center, where a negative mass singularity is present. Also a dust source that gravitates but does not interact with the particles in the accretion disk can be interpreted as a rotating cloud of dark matter. Our aim here is to investigate whether such fluid can be used to model an astrophysical supermassive compact object and if such a hypothetical source can be distinguished from a black hole via the emission spectrum of its accretion disk.

The article is organized as follows. In Section~\ref{metric}, we briefly discuss stationary axially symmetric solutions of Einstein's field equations and in particular rigidly rotating dust. In Section~\ref{accr}, we describe the properties of accretion disks in such a solution and compare the luminosity spectrum emitted by the disk with that coming from a black hole. In Section~\ref{K}, we simulate the emission of the iron K$\alpha$ line from such accretion disks and discuss whether such lines can be detected in observations of supermassive compact objects. Section~\ref{disc} is devoted to discussing the results and future directions. Throughout the article we make use of natural units setting $G=c=1$.

\section{Rotating dust}\label{metric}

The metric for a stationary, axially symmetric rigidly rotating dust source was found by Van Stockum
(see \cite{vanstockum,islam})
and we can write it in cylindrical coordinates $\{t,\rho,\phi,z\}$ as
\be
ds^2=-dt^2+2\eta dtd\phi+e^\mu(d\rho^2+dz^2)+(\rho^2-\eta^2)d\phi^2 \; ,
\ee
where $\eta(\rho,z)$ and $\mu(\rho,z)$ are the metric functions to be determined via Einstein's equations.
If we define a function $\xi(\rho,z)$ from
\be
\eta(\rho,z)=\rho\xi_{,\rho} \; ,
\ee
then the problem reduces to solving
\begin{eqnarray} \label{laplace}
\nabla^2\xi&=&0 \; , \\ \label{mur}
\mu_{,\rho}&=&\frac{\eta_{,z}^2-\eta_{,\rho}^2}{2\rho} \; , \\ \label{muz}
\mu_{,z}&=&-\frac{\eta_{,z}\eta_{,\rho}}{\rho} \; ,
\end{eqnarray}
where $\nabla^2$ indicates the Laplace operator in the flat three-dimensional space.
Once a solution of equation (\ref{laplace}) is given, equations (\ref{mur}) and (\ref{muz}) are immediately integrated by quadrature.
Therefore the whole problem of finding a solution of Einstein's equations describing stationary, axially symmetric, dust reduces to the problem of solving Laplace's equation in ordinary flat three-dimensional space. Note that the same is true for static, vacuum, axially symmetric solutions, and therefore there exist a one to one correspondence between static, vacuum, axially symmetric solutions and stationary, axially symmetric rigidly rotating dust.
Einstein's equations for dust take the form
\be
R_{ij}-\frac{1}{2}g_{ij}R=8\pi\epsilon u_i u_j \; ,
\ee
where $u_i$ is the four-velocity (given by $u^i=\delta^i_0$) and $\epsilon(\rho,z)$ is the energy density of the dust cloud and it is given by
\be\label{density}
8\pi\epsilon=\frac{e^{-\mu}}{\rho^2}(\eta_{,\rho}^2+\eta_{,z}^2) \; .
\ee

Since we are considering stationary, axially symmetric configurations we will look for a solution of equation (\ref{laplace}) that does not depend on $t$ and on the angular variable $\phi$. Also, the fact that the metric is stationary and axially symmetric implies that we have the usual commuting Killing vectors
$T^\mu=\delta^\mu_t$ and $\Psi^\mu=\delta^\mu_\phi$.
The four-velocity is then given by $u^\mu=T^\mu+\Omega_0 \Psi^\mu$, where $\Omega_0$ is the (constant) angular velocity.

\subsection{Bonnor's solution}
The simplest solution of equation (\ref{laplace}) can be obtained taking
\be
\xi=\frac{2h}{r} \; ,
\ee
where we have set $r=\sqrt{\rho^2+z^2}$ and $h$ is an arbitrary constant.
This is the space-time first investigated by Bonnor in \cite{Bonnor}.
For the metric functions we get
\begin{eqnarray}
\eta&=&-\frac{2h\rho^2}{r^3} \; , \\
\mu&=&\frac{h^2\rho^2(\rho^2-8z^2)}{2r^8} \; ,
\end{eqnarray}
and the corresponding energy density is given by
\be
8\pi \epsilon=\frac{4h^2e^{-\mu}}{r^8}(4z^2+\rho^2) \; .
\ee
For this model it is relatively easy to evaluate $\mu$ and $\eta$ and at spatial infinity the metric tends to become Minkowski in all directions.
The energy density is always positive and it drops rapidly as $r$ increases but on the other hand it presents a singularity at $r=0$.
The total mass of the system is zero due to a distributional infinite negative mass located at the singularity, which balances the positive mass and makes the metric somehow unphysical.
Nevertheless, thinking about an astrophysical source, the presence of a singularity at the center indicates a regime where classical General Relativity doesn't hold anymore.
In fact, diverging quantities typically suggest a breakdown of the model and in particular singularities in GR are generally associated with a failure of the classical theory to describe strong gravitational fields over short distances. In order to have a complete description of the behaviour of gravity in these extreme conditions one would need to use a theory of quantum gravity. However, despite the lack of a viable quantum gravity theory at the moment, there exist several indications of the possible corrections that such a theory would induce on classical models (see Ref.~\cite{universe} and reference therein). Therefore it is reasonable to argue that the singularity at the center of the Bonnor space-time would be resolved by a theory of quantum gravity and we can assume that the present solution will be valid from a certain radius $r_0$ outwards, while the behaviour for $r<r_0$ can not be properly captured by the classical solution. We can further speculate that the resolution of the singularity may imply the existence of an exotic compact object at the center. Exotic compact objects have been proposed for decades also within classical GR. The size and properties of such objects vary from model to model. For example, objects like quark stars \cite{quark} and boson stars \cite{boson} are slightly bigger than the Schwarzschild radius for a black hole of the same mass, while gravastars \cite{grava} for example have a radius of the order of the Schwarzschild radius. Other, more compact, objects have been suggested as well (see for example Ref.~\cite{planck}) and they may be intrinsically quantum in nature.

In this case, we assume that we can neglect the finite amount of matter enclosed between the center and $r_0$ and evaluate the total positive mass of the system as the amount of matter contained between $r_0$ and infinity.
We find
\be\label{M}
M_0=\frac{3\pi h^2}{16r_0^3} \; .
\ee
Note that as $r_0$ goes to zero the mass diverges, while the total mass for the system is zero. This suggests that the central singularity is acting as a distribution of infinite negative mass that balances the positive mass of the space-time.
We shall mention here that another way to deal with the presence of the singularity is by constructing a solution describing a rigidly rotating dust configuration accompanied by a rigidly rotating thin disk located on the equatorial plane
(see for example Ref.~\cite{Neu} for a relativistic rigidly rotating disk of dust in vacuum).
The metric then results from the matching of three parts, two rigidly rotating dust space-times one above and one below the equatorial plane, and the disk. The matter distribution on the disk is then given by jump conditions for the second fundamental form across the disk. Such a model effectively removes the plane where the singularity is located and replaces it with another matter source.
The properties of this kind of models are beyond the scope of the present article and will be investigated elsewhere.

\subsection{Rotating dust as dark matter}

As we mentioned before our aim is to investigate the possibility that relativistic rotating dust could be used to describe dark matter clouds at galactic centers, possibly without invoking the presence of supermassive black holes.
Dark matter clouds in galaxies can be described by Newtonian density profiles that give a phenomenological account of the missing matter distribution that is necessary to explain the velocity dispersion of stars in the outer regions of galaxies.

One of the most widely used density profiles is the Navarro-Frenk-White (NFW) model for which the density takes the form
\be
\epsilon=\frac{\epsilon_0}{\frac{r}{r_s}\left(1+\frac{r}{r_s}\right)^2},
\ee
where $\epsilon_0$ is a constant describing the characteristic density of the model and $r_s$ is a parameter related to the scale \cite{nfw}.
The NFW profile is a variation of the Jaffe profile
\cite{jaffe}, given by
\be
\epsilon=\frac{\epsilon_0}{\left(\frac{r}{r_s}\right)^2\left(1+\frac{r}{r_s}\right)^2},
\ee
which was originally derived by observing the brightness of spiral galaxies.
Other density profiles that can be considered are the pseudo-isothermal sphere
\cite{iso}, given by
\be
\epsilon=\frac{\epsilon_0}{1+\left(\frac{r}{r_s}\right)^2},
\ee
or the Universal Rotation Curve profile
\cite{urc},
given by
\be
\epsilon=\frac{\epsilon_0}{\left(1+\frac{r}{r_s}\right)\left(1+\left(\frac{r}{r_s}\right)^2\right)}.
\ee

As said, these profiles are constructed in order to explain the observed behaviour of galaxies and galaxy clusters at large scales and they need not provide a good description for dark matter near the center. In fact some profiles, like the NFW for example, present cusps or divergencies at the center.

The density profile obtained from the Bonnor solution is also not well behaved at the center (as it goes like $1/r^6$ on the equatorial plane, near the center), presenting an even steeper divergence with respect to the other profiles. On the other hand the Bonnor profile differentiates itself from the standard profiles mentioned above in the fact that it is axially symmetric and it includes the effects of rotation (notice that both features may be testable in principle). Of course this need not be the best description for dark matter near galactic centers as there may be more realistic models, that possibly take into consideration a transition from a Newtonian profile at large distances to a relativistic profile near $r=0$. However, the main point that we want to stress here is that relativistic effects due to the high density reached towards the inner region of the galaxy may play an important role and affect the dynamics of accreting gas. In this case a simple analytical relativistic model like the Bonnor space-time does provide indications on the general behaviour of ordinary matter moving inside a dense but pressureless fluid and allows for comparison with the motion of particles in a vacuum black hole space-time.

\subsection{Other solutions}
As we have said, to every solution of Laplace equation corresponds a rigidly rotating dust space-time. Therefore other matter sources can be found by solving equation (\ref{laplace}) via separation of variables with $\xi$ not depending on $\phi$. If we consider the simplest non trivial solution given by
\be\label{2}
\xi(\rho,z)=J_0(\rho)e^{-\alpha|z|} \; ,
\ee
where $\alpha$ is a positive constant and $J_0(\rho)$ is the zero order Bessel function of the first kind, we can obtain a density profile that does not diverge anywhere.
This solution was studied in
Ref.~\cite{wang}
in connection to the possible appearance of jets powered by a non vanishing density gradient along the axis.
In this case the expressions for $\eta$ and $\mu$ become rather complicated and involve Bessel functions.
The energy density for this model can be calculated from equation (\ref{density}) and it is finite along the central axis while again it decreases rapidly away from the axis.
Note however that the above solution is not asymptotically flat.
It is known, for example, that a spherically symmetric dust cloud, matched to a vacuum asymptotically flat exterior must necessarily collapse. Similarly in order to prevent a rigidly rotating dust solution from collapsing one has to introduce some other effects. In the case of the Bonnor solution it is the negative mass distribution at the singularity.
For rotating dust solutions with a regular density profiles such as the one given by equation (\ref{2}), in order to have a stationary configuration one must allow for some energy distribution at infinity. In this case the space-time will be asymptotically anti-DeSitter, to balance the positive energy of the dust profile that would cause the whole configuration to collapse.

\section{Accretion disks}\label{accr}

In this section we study the motion of test particles moving on circular geodesics in the equatorial plane of the Bonnor space-time. These particles represent particles in the accretion disk of the space-time, while the dust particles of the source represent a dark matter cloud that extends throughout the galaxy and has higher density towards the center. So we model a dark matter cloud near the center of a galaxy by assuming that we can have circular geodesics inside the source, and that test particles will move on such geodesics unaffected. Dark matter has detectable gravitational effects but does not interact, or interacts very weakly, with ordinary matter. For this reason we consider dust as a viable theoretical model for dark matter.

If we pass to spherical coordinates $\{r,\theta\}$ via the transformation $z=r\cos\theta$ the metric takes the form
\bea \nonumber
ds^2&=&-dt^2+ 2 \eta dt d\phi^2+e^\mu (dr^2+r^2 d\theta^2)+ \\
&+&(r^2\sin^2\theta -\eta^2)d\phi^2 \; ,
\eea
where now $\eta$ and $\mu$ are functions of $r$ and $\theta$.
For an observer at infinity, the Bonnor space-time is equivalent to that of a spinning massless particle located at the origin. In order to describe test particles in the accretion disk, we will restrict our analysis to the equatorial plane thus taking $\theta\simeq \pi/2$.
Given the existence of the two killing vectors associated with time translations and spatial rotations we can define two conserved quantities, namely the energy $E$ and the angular momentum $L$ for particles in the accretion disk. Then we can express the effective potential for a test particle in the equatorial plane of the the Bonnor space-time as
\bea \nonumber
V_{\rm Bonn}^{\rm eff}&=&1-\frac{E^2g_{\phi\phi}+2ELg_{t\phi}+L^2g_{tt}}{g_{t\phi}^2-g_{tt}g_{\phi\phi}}=\\
&=&1-E^2+\frac{L^2}{r^2}+\frac{4ELh}{r^3}+\frac{4E^2h^2}{r^4} \; .
\eea

Note the absence of the term in $1/r$ as opposed to the Schwarzschild or Kerr case. This is due to the fact that the only source of gravity for the Bonnor space-time is angular momentum (the total mass being zero).
Also note that, in order to have bound motion in the equatorial plane, we must take $h<0$. This can be understood again as the result of the absence of an attractive positive mass, thus implying that all the gravitational effects are due to the angular momentum of the source.
For comparison the effective potentials for Schwarzschild and Kerr are given by
\bea
V_{\rm Schw}^{\rm eff}&=& 1-E^2-\frac{2M}{r}+\frac{L^2}{r^2}-\frac{2ML^2}{r^3} \; , \\
V_{\rm Kerr}^{\rm eff}&=& 1-E^2-\frac{2M}{r}+\frac{L^2+a^2(1-E^2)}{r^2}-\frac{2M(aE-L)^2}{r^3} \; .
\eea

\begin{figure}[ht]
\centering
\begin{minipage}{.48\textwidth}
\centering
\includegraphics[scale=0.6]{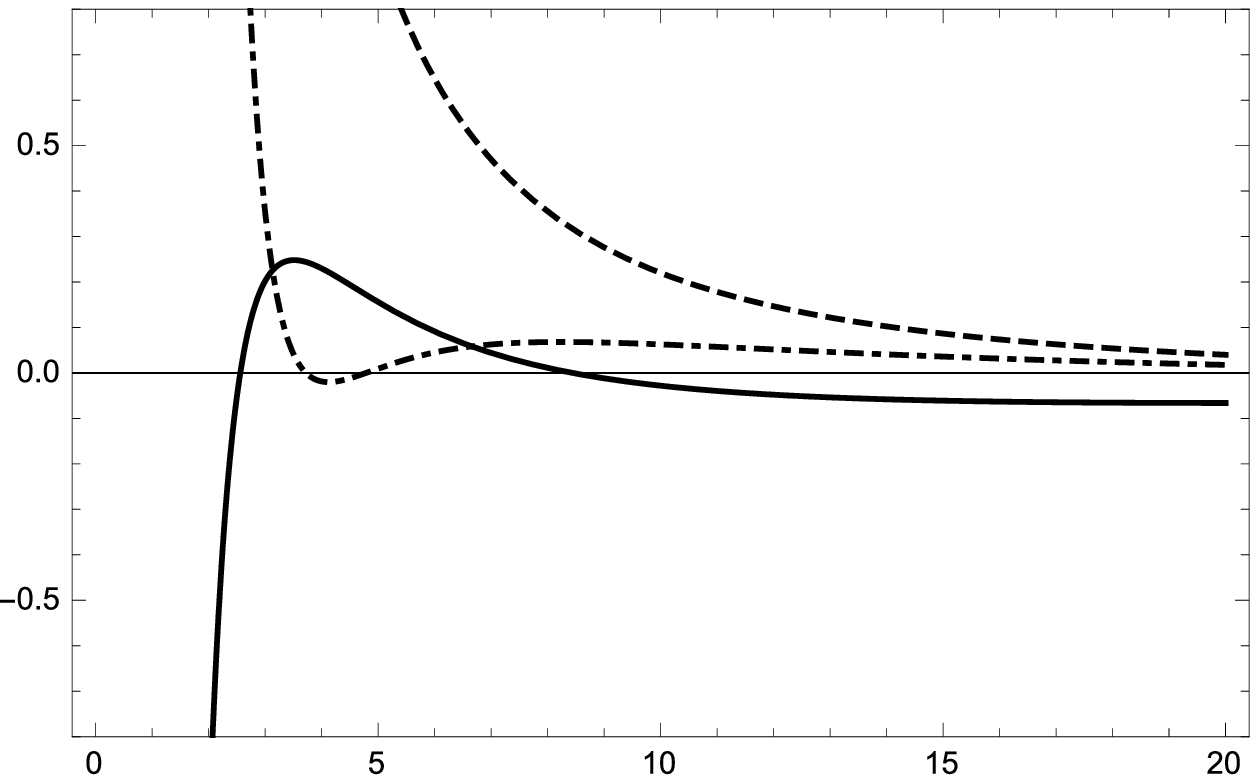}
\put(3,2){$r$}
\put(-224,124){$V^{\rm eff}$}
\end{minipage}
\hfill
\begin{minipage}{.48\textwidth}
\centering
\includegraphics[scale=0.6]{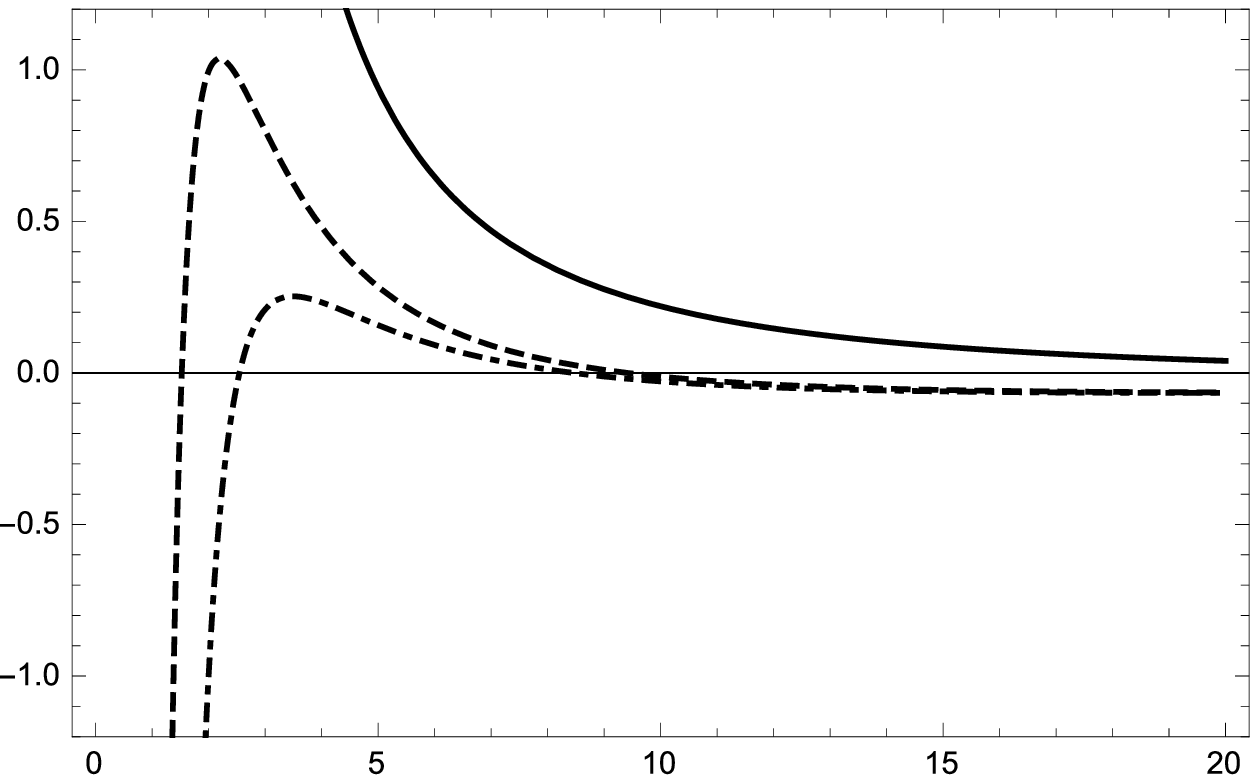}
\put(3,2){$r$}
\put(-224,126){$V^{\rm eff}$}
\end{minipage}
\caption{Top panel: Comparison of the effective potential between two models of the Bonnor solution (dashed line with $h=-0.000214$ and dot-dashed line with $h=-10$) with the Schwarzschild case (continuous line) with $M=1$. Chosen values for energy and angular momentum are $E=1.01$ and $L=\sqrt{24}$.
Bottom panel: Comparison of the effective potential between the Bonnor model with $h=-0.000214$ (continuous line) and Kerr solutions with $M=1$ and angular momentum parameter $a=0.01$ (dot-dashed line) and $a=0.9$ (dashed line). Chosen values for energy and angular momentum are $E=1.01$, $L=\sqrt{24}$.}
\label{fig1}
\end{figure}

From the condition for circular motion, given by $V_{\rm eff}(r)=V_{\rm{eff},r}(r)=0$, we see that
\bea
E&=& -\frac{g_{tt}+g_{t\phi}\Omega}{\sqrt{-g_{tt}-2g_{t\phi}\Omega-g_{\phi\phi}\Omega^2}}=\frac{r^2}{\sqrt{r^4-4h^2}} \; , \\
L&=& \frac{g_{t\phi}+g_{\phi\phi}\Omega}{\sqrt{-g_{tt}-2g_{t\phi}\Omega-g_{\phi\phi}\Omega^2}}=\frac{4|h|r}{\sqrt{r^4-4h^2}} \; ,
\eea
where $\Omega$ is the angular velocity of the particles in the accretion disk and it is given by
\be
\Omega=\frac{d\phi}{dt}=\frac{\sqrt{g_{t\phi ,r}^2-g_{tt,r}g_{\phi\phi ,r}}-g_{t\phi , r}}{g_{\phi\phi ,r}}=\frac{2|h|r}{r^4+4h^2} \; .
\ee

Circular orbits can exist only if $g_{tt}+2g_{t\phi}\Omega+g_{\phi\phi}\Omega^2\leq 0$, thus only for $\Omega_-<\Omega<\Omega_+$, where
\be
\Omega_\pm=\omega\pm\sqrt{\omega^2-\frac{g_{tt}}{g_{\phi\phi}}} \; ,
\ee
and $\omega=-g_{t\phi}/g_{\phi\phi}$ is the frame dragging frequency of the space-time. The limiting case for which $\Omega=\Omega_\pm$ defines the photon capture spheres of the space-time.
It is easy to see from Figure \ref{fig1} that as particles get closer to the center the behaviour of the effective potential differs drastically between the Bonnor model and the Schwarzschild or Kerr black holes. This suggests that the emission spectrum of such accretion disks will also be considerably different.

\begin{figure}[t]
\centering
\begin{minipage}{.48\textwidth}
\centering
\includegraphics[scale=0.6]{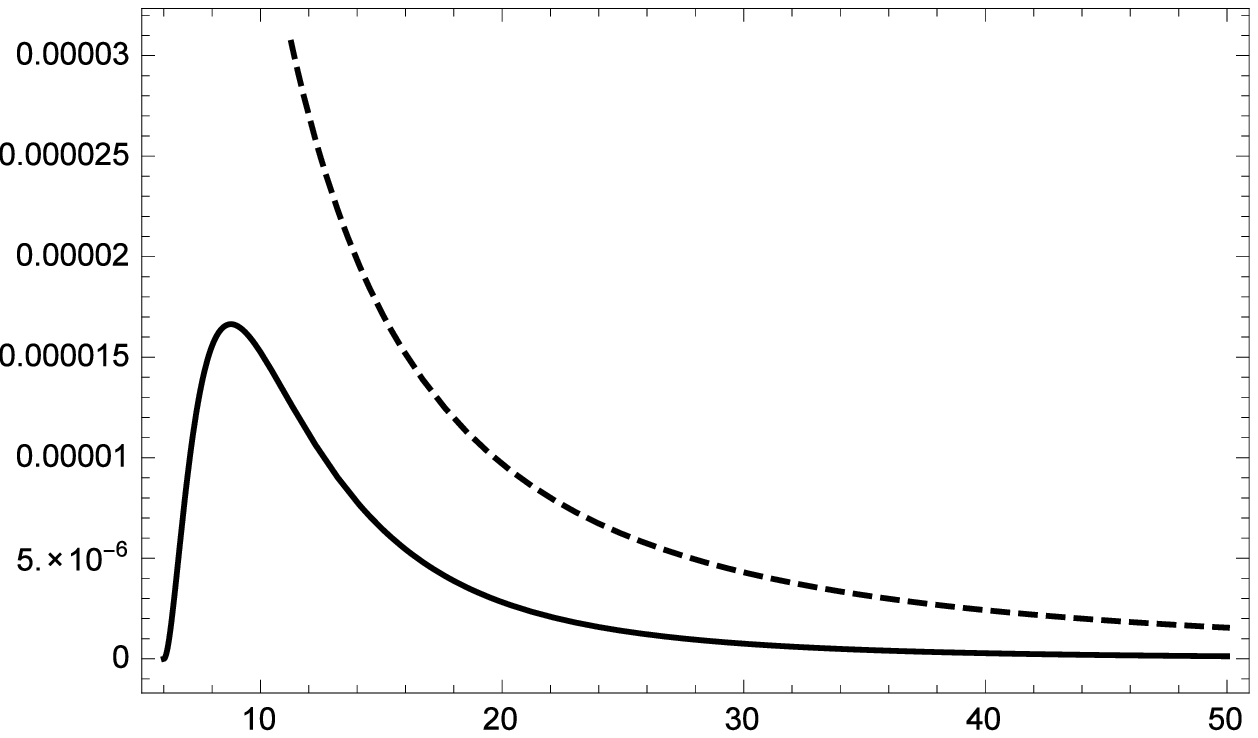}
\put(-220,90){$f/\dot{m}$}
\put(3,3){$r/M$}
\end{minipage}
\hfill
\begin{minipage}{.48\textwidth}
\centering
\includegraphics[scale=0.6]{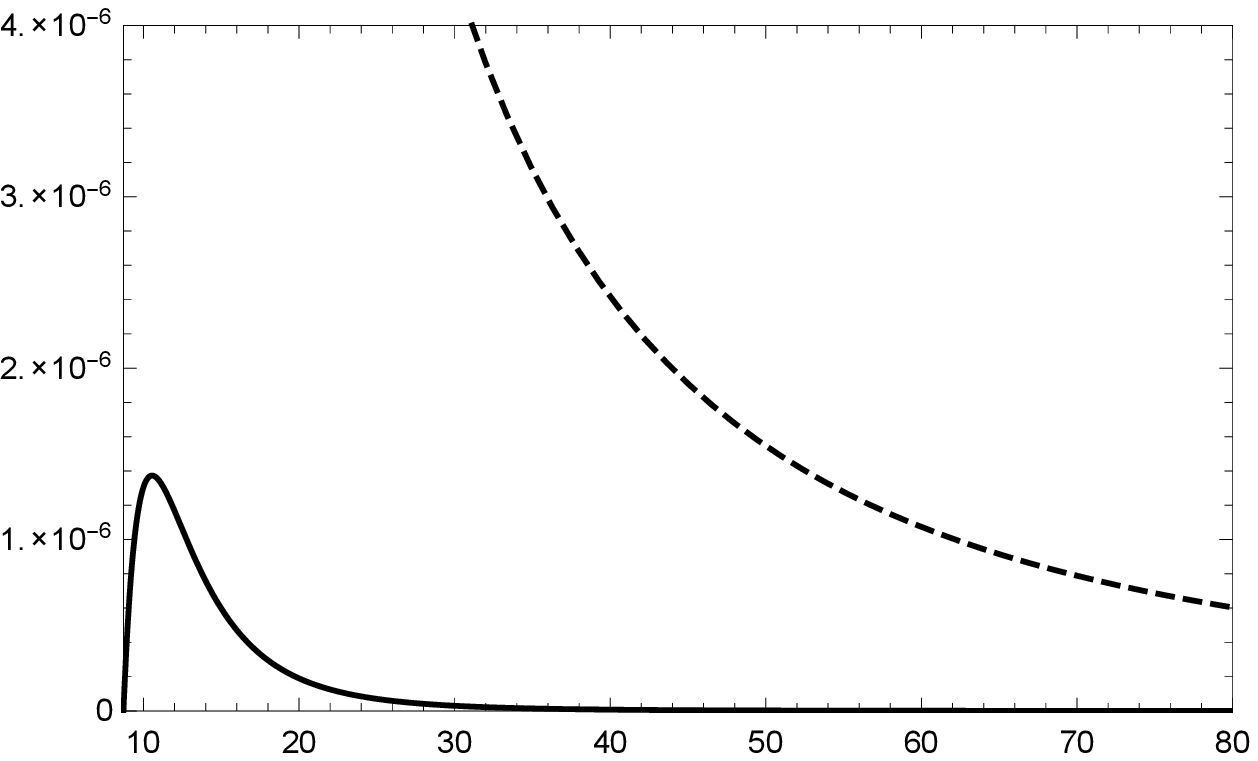}
\put(-220,80){$f/\dot{m}$}
\put(3,3){$r/M$}
\end{minipage}
\caption{Top panel: Comparison between the luminosity flux from accretion disks around a Schwarzschild black hole (continuous line) and the luminosity flux from accretion disks inside the Bonnor solution (dashed line) with $h=-0.000214$. The value of the constant $h$ has been chosen in order to have the mass of the Bonnor solution between $r_0$ and infinity (given by equation (\ref{M})) equal to the Schwarzschild mass $M$ as measured by observers at infinity. For this particular plot the chosen value of $r_0$ was $r_0=0,003M$.
Bottom panel: Comparison of the luminosity flux in the Bonnor space-time (dashed line) with a Kerr black hole with $a=0.9$ (continuous line). The value of the ISCO for such a black hole is $r\simeq 8.7M$. Note that the luminosity depends on the location of the ISCO and therefore on the value of $a$, however it is generally lower than the Bonnor case. For $h$, $M$ and $r_0$ the values chosen are the same as the ones in the top panel.
 }
\label{fig2}
\end{figure}

The next step is to compare the luminosity flux of the accretion disk in the Bonnor space-time with that of a black hole.
The luminosity flux per unit accretion mass is given by
\be
\frac{f(r)}{\dot{m}}=-\frac{\omega_{,r}}{\sqrt{-g}(E-\omega L)^2}\int_{r_{\rm ISCO}}^r(E-\omega L)L_{,r}dr \; ,
\ee
where $\dot{m}$ is the mass accretion rate onto the central object and $r_{\rm ISCO}$ is the radius of the innermost stable circular orbit (ISCO).
Usually the ISCO, which marks the boundary of the accretion disk, is defined by the condition that $V_{{\rm eff},rr}=0$ and for Schwarzschild, for example, it is located at three times the horizon radius. As for the Bonnor space-time it is easy to verify that there is no ISCO, meaning that particles are allowed to move in circular orbits all the way to the central singularity. This is similar to the behaviour found in other space-times describing non vacuum interiors
(see for example \cite{BM})
and therefore it is natural to take another parameter as the  boundary of circular orbits. For the Bonnor space-time we take $r_0=r_{\rm ISCO}$ and we evaluate the luminosity flux per unit accretion as
\bea \nonumber
\frac{f(r)}{\dot{m}}&=&\frac{2\sqrt{2}|h|(4h^2-3r^2)}{e^{h^2/2r^4}(4h^2+r^4)\sqrt{r^4-4h^2}} \\
&&\left(\tan^{-1}\sqrt{r}-\frac{1}{2}\ln\left|\frac{\sqrt{r}-1}{\sqrt{r}+1}\right|+C(r_0)\right) \; ,
\eea

where $C(r_0)$ in an integration constant that depends on the choice of $r_0$. The plot of the luminosity is in Figure \ref{fig2}. As expected, given the presence of the singularity, the total luminosity flux for the Bonnor space-time is diverging. Making the assumption that the singularity is removed we can still notice that the total luminosity given away by such an object would be much greater than the luminosity emitted by the accretion disk surrounding a Schwarzschild black hole. This is in agreement with what was found in
\cite{obs}
and suggests the possibility that very luminous sources may be less massive exotic compact objects rather than very massive black holes.

\section{Iron K$\alpha$ line}\label{K}

Within the disk-corona model, a black hole is surrounded by a geometrically thin and optically thick accretion disk that emits like a black body locally and a multi-color black body when integrated radially. The corona is a hot ($\sim 100$~keV), usually optically thin, electron cloud enshrouding the disk. Due to inverse Compton scattering of some thermal photons of the disk off the hot electrons in the corona, the latter becomes an X-ray source with a power-law spectrum. A fraction of X-ray photons can illuminate back the accretion disk, producing a reflection component with some fluorescence emission lines. The most prominent feature in the reflection spectrum is the iron K$\alpha$ line. In the case of neutral iron, this line is at 6.4~keV, but it can shift up to 6.97~keV in the case of H-like iron ions (which can be the case in the accretion disk of a stellar-mass black hole). For a review, see e.g. Ref.~\cite{fabian}.

The study of the shape of the iron line can be a powerful tool to probe the space-time metric around black hole candidates and test the nature of these compact objects~\cite{i1,i2,i3,i4}. The exact shape of the iron line is determined by the metric of the space-time, the geometry of the emitting region in the accretion disk, the intensity profile, and the inclination angle of the disk with respect to the line of sight of the distant observer. The choice of the intensity profile is crucial in the final measurement, but current studies usually assume a simple power law, namely the local intensity is $I_{\rm e} \propto 1/r^q$ where the emissivity index $q$ is a free parameter to be determined by the fit. A slightly more sophisticated model is a broken power-law, where $I_{\rm e} \propto 1/r^{q_1}$ for $r < r_{\rm b}$, $I_{\rm e} \propto 1/r^{q_2}$ for $r > r_{\rm b}$, and there are three free parameters, the two emissivity indexes $q_1$ and $q_2$ and the breaking radius $r_{\rm b}$.

\begin{figure}[hh]
\centering
\begin{minipage}{.48\textwidth}
\centering
\includegraphics[scale=0.65]{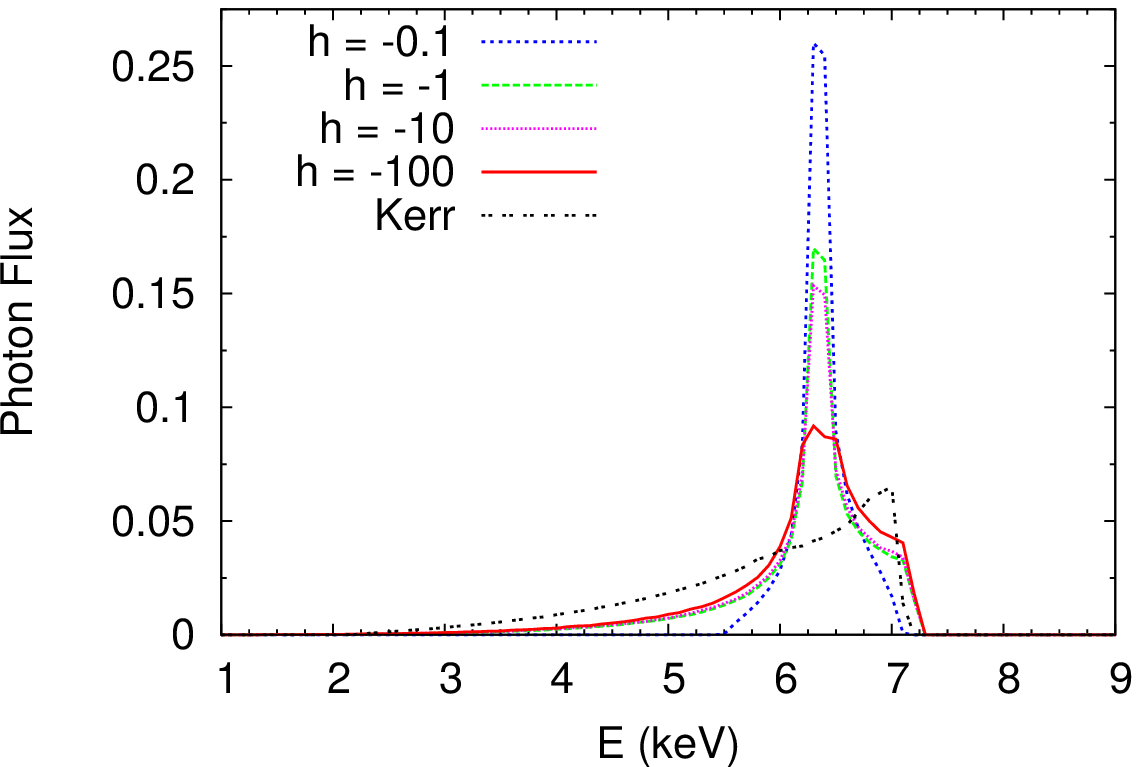}
\end{minipage}
\hfill
\begin{minipage}{.48\textwidth}
\centering
\includegraphics[scale=0.65]{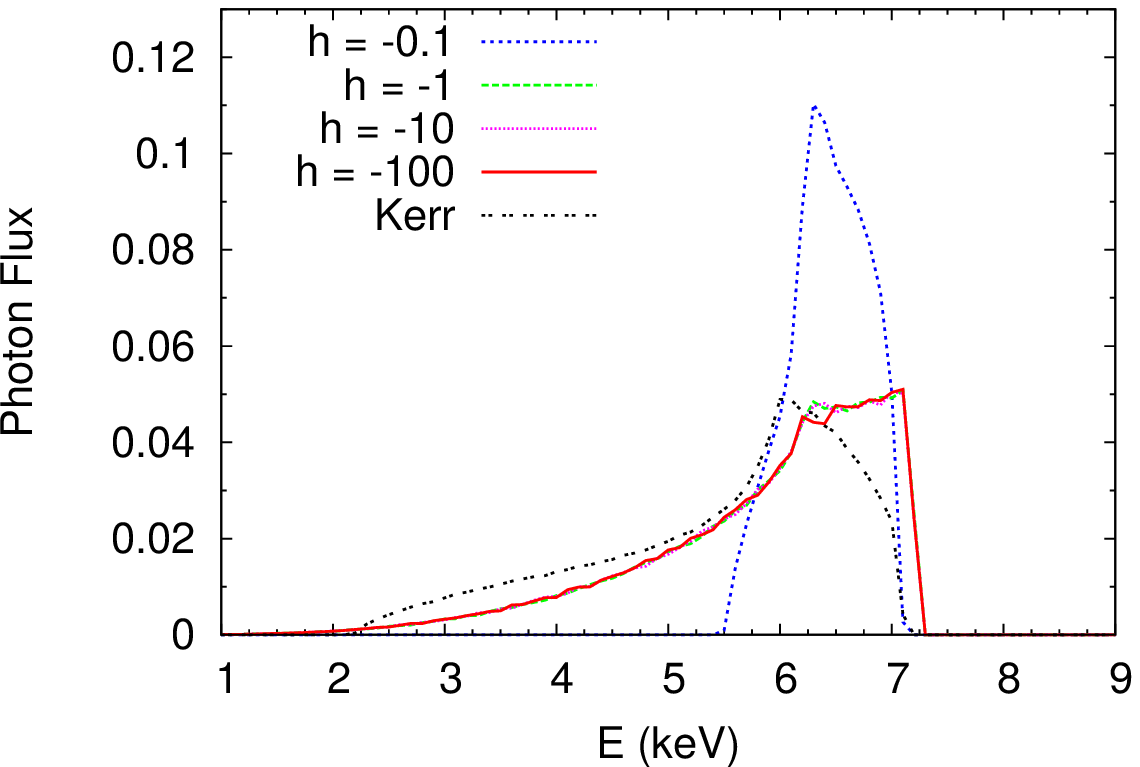}
\end{minipage}
\vspace{0.6cm}
\caption{Shape of the iron K$\alpha$ line expected in the reflection spectrum of an accretion disk in the Bonnor space-time for different values of the constant $h$. These lines have been computed assuming the viewing angle $i = 45^\circ$, the inner edge of the disk $r_{\rm in} = 1$, and the intensity profile $I_{\rm e} \propto 1/r^q$, where the emissivity index is $q = 3$ (top panel) and 5 (bottom panel).
For comparison the corresponding line in the Kerr space-time with $a=0.7$ is plotted as well (see text for more details).}
\label{fig3}
\end{figure}

The calculations of the shape of the iron line in a generic stationary, axisymmetric, and asymptotically flat space-time have been already extensively discussed in the literature. In our case, we use the code described in~\cite{code1,code2}. The accretion disk is described by the Novikov-Thorne model, where the disk is in the equatorial plane perpendicular to the object's spin and the particles of the gas follow nearly geodesic circular orbit in the equatorial plane. The inner edge of the disk is at the ISCO radius. In the Bonnor space-time there is no ISCO and therefore we set the inner edge at some arbitrary radius $r_{\rm in}$. Some examples of iron K$\alpha$ line in the reflection spectrum of an accretion disk in Bonnor space-times are shown in Fig.~\ref{fig3}, where $r_{\rm in} = 1$. An accretion disk only exist for $h < 0$ and the iron line becomes broader as $h$ decreases. In both panels in Fig.~\ref{fig3}, we assume the intensity profile $I_{\rm e} \propto 1/r^q$, where $q=3$ in the top panel and $q = 5$ in the bottom panel. As we can see, the intensity profile plays an important role in the final shape of the line.


In order to understand whether current X-ray observations of AGN can already rule out, or constrain, the possibility that the metric around the supermassive compact objects in galactic nuclei can be described by the Bonnor solution discussed in the present paper, we follow the strategy already employed in Refs.~\cite{x1,x2,x3}. The reflection spectra in current X-ray data are commonly fitted with Kerr models and there is no tension between predictions and observational data. This means that, even if the space-time metric around AGN were not described by the Kerr solution, current data would not be able to unambiguously identify deviations from the Kerr space-time. We thus simulate observations with a current X-ray mission employing iron lines computed in the Bonnor metric and we fit the simulations with a Kerr model. If the fit is acceptable, we conclude that current data cannot exclude the possibility that Bonnor metrics describe the space-time around AGN. In the opposite case, if the fit is bad, we argue that current data can rule out the Bonnor metric of the simulation.

As theoretical model for the simulations, we employ a simple power-law with photon index $\Gamma = 2$ to describe the spectrum of the hot corona and a single iron K$\alpha$ line to describe the reflection spectrum of the disk. We assume the typical parameters of a bright AGN with a strong iron line. The flux in the energy range 0.7-10~keV is about $2 \cdot 10^{-10}$~erg/s/cm$^2$ and the iron line equivalent with is about 200~eV. We convolve this spectrum with the response of the instrument in order to get the simulated observation. The simulations are done considering the detector XIS0 on board of the X-ray mission Suzaku and assuming that the expose time is 1~Ms.

\begin{widetext}

\begin{figure}[ht]
\centering
\includegraphics[scale=0.5,angle=270]{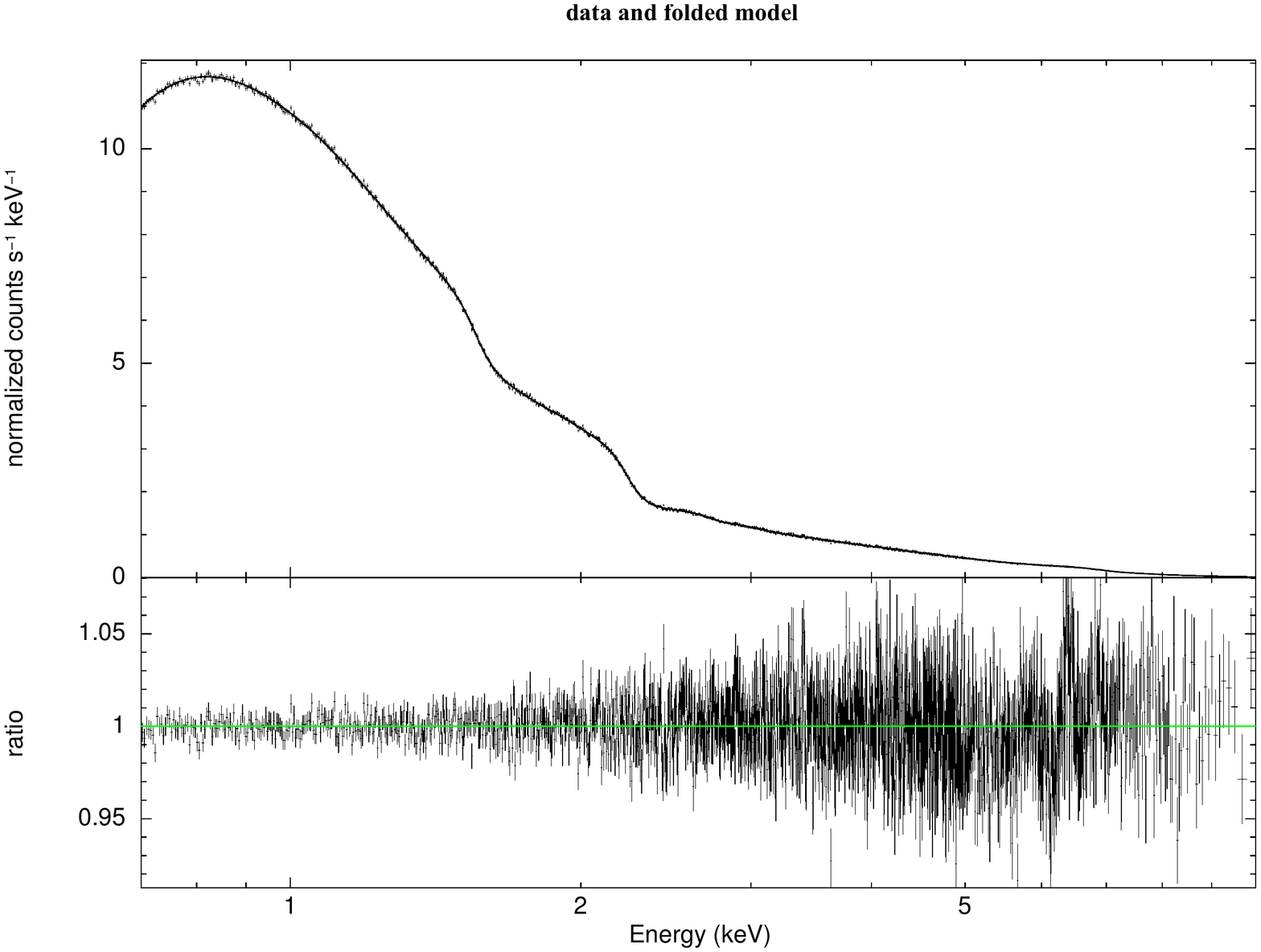}
\caption{Top panel: simulated data and best fit of the observed spectrum assuming that the source is a bright AGN. Bottom panel: ratio between the simulated data and the best fit. The spectrum of the source is computed assuming the Bonnor metric with $h = -0.1$. The viewing angle is $i=45^\circ$ and the exposure time is $\tau = 1$~Ms. The minimum of the reduced $\chi^2$ is about 1.07. See the text for more details.}
\label{fig4}
\end{figure}


\begin{figure}[ht]
\centering
\includegraphics[scale=0.5,angle=270]{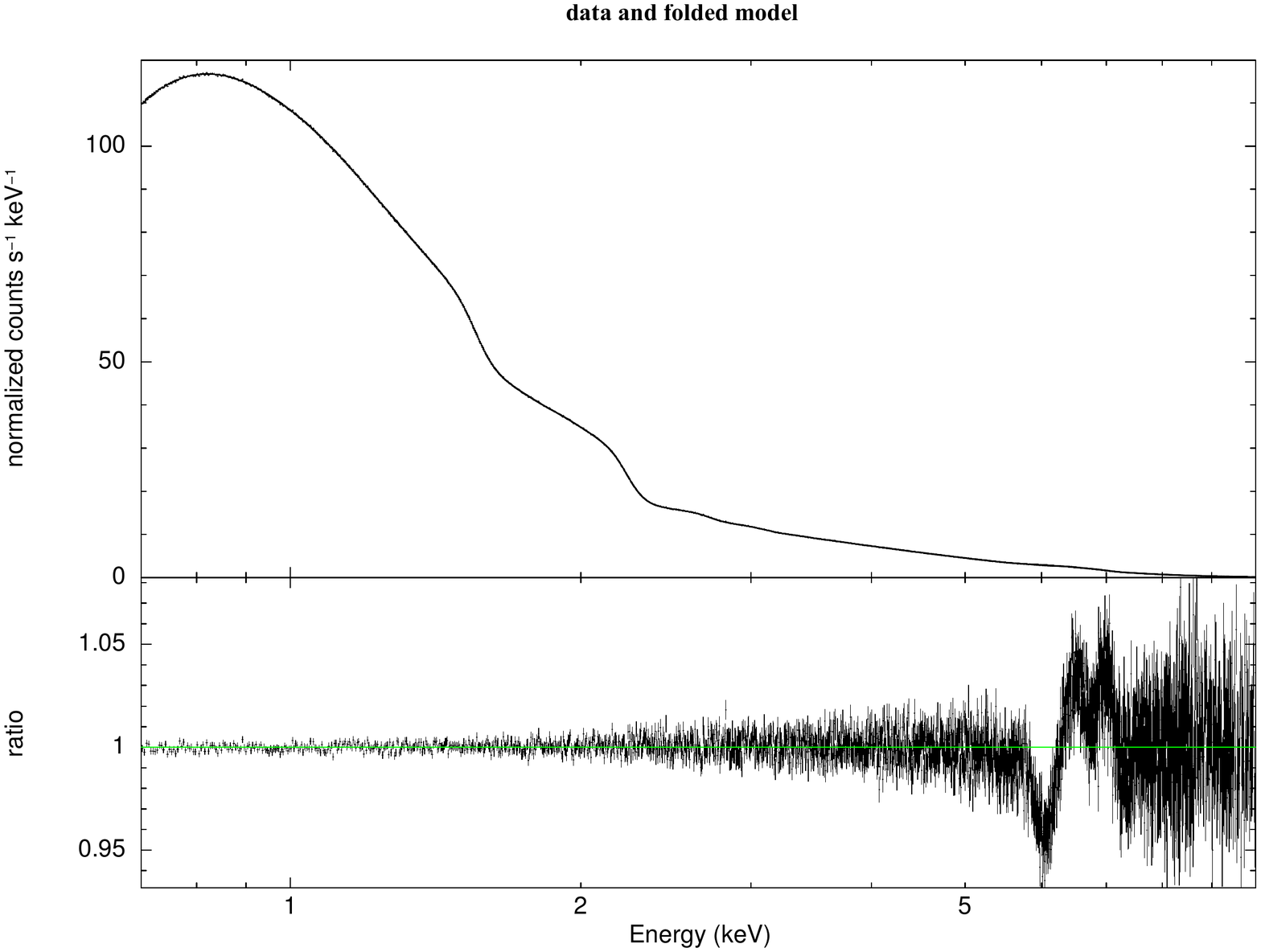}
\caption{As in Fig.~\ref{fig4}, but assuming that the source is a bright black hole binary; that is, the energy flux is an order of magnitude higher. We clearly see unresolved features at 6-7~keV. The minimum of the reduced $\chi^2$ is about 1.98. See the text for more details.}
\label{fig5}
\end{figure}

\end{widetext}

Our simulations show that we can always obtain good fits for any value of $h$. The crucial point is the emissivity index $q$. At the moment, we do to know the exact geometry of the corona, and therefore we cannot predict the exact intensity profile. For example, if we employ $q=3$ in the simulations, our iron line cannot be fitted with a Kerr model. However, if we use $q=5$, we alway find acceptable fits. An example is shown in Fig.~\ref{fig4}, where we see (bottom panel) that the ratio between simulated data and the Kerr model is always close to 1. We can thus assert that current observational facilities are likely unable to test the Bonnor metric in AGN. The limitation is in the current photon count. For example, if we simulate the observation of a stellar-mass black hole, namely we assume that the energy flux is $2 \cdot 10^{-9}$~erg/s/cm$^2$, we see the difference, as shown in Fig.~\ref{fig5}. Future X-ray missions, with a larger effective area, can thus test the Bonnor metric in AGN.

\section{Discussion}\label{disc}

Strong gravitational fields generated by axially symmetric rotating matter sources can be of importance in astrophysics, especially when it comes to understanding the supermassive compact objects that dwell at the center of galaxies. In the present article, we have considered a standard way to test the assumption that such objects must be Kerr black holes by comparing the properties of accretion disks around black holes with those of some space-time metric that does not describe a black hole.

Our main assumption is that the galactic dark matter halos require a relativistic description towards the galactic center and that the Bonnor space-time (describing rigidly rotating dust) may provide a good first approximation for a relativistic model of dark matter.
The most commonly used models for dark matter, like for example the Jaffe density profile \cite{jaffe} or the NFW profile \cite{nfw}, present a divergence of the density at the center while at the same time being purely Newtonian. It is then natural to assume that relativistic effects will become important in regions where the density becomes high. These regions are also the centers of galaxies where we believe that supermassive black holes reside. Therefore we have considered a simple relativistic space-time describing rotating dust and suggested the possibility that this kind of matter distributions, together with the effects of general relativity, may be sufficient to explain the behaviour of accretion disks near the galactic centers, possibly without invoking the presence of supermassive black holes. To this aim we have constructed a simple observational test that in the near future could be used to check the validity of such models.

We have shown that the effective potential for ordinary massive particles moving in the accretion disk of the Bonnor solution is considerably different with that of black hole solutions.
As a consequence, the luminosity spectrum of the accretion disk in such a space-time will also differ substantially from the spectrum of accretion disks around black holes. The shape of the iron K$\alpha$ line in the reflection spectrum is also substantially different and it can be used to test this scenario with future X-ray missions. Future observations of the shadow of the supermassive compact object at the galactic center will soon allow us to test the assumption that such object must be a black hole.

\end{document}